\documentclass[aps,prl,twocolumn,showpacs,reprint]{revtex4-1}

\usepackage{wrapfig}
\usepackage[]{graphicx,xcolor}
\usepackage{tabularx}
\usepackage{booktabs}
\usepackage{textcomp}
\usepackage{amsmath}
\usepackage{amssymb}
\usepackage[]{graphics}
\usepackage{amssymb}
\usepackage{amsmath}
\usepackage{color}
\usepackage{ifpdf}
\newcommand{\executeiffilenewer}[3]{%
\ifnum\pdfstrcmp{\pdffilemoddate{#1}}%
{\pdffilemoddate{#2}} > 0 {\immediate\write18{#3}}\fi}

\ifpdf\usepackage{epstopdf}\fi
\newcommand{\si}{SiO$_2$ }
\newcommand{\va}{VO$_2$ }

\begin{document}

\title{Near field thermal memory device}

\author{S. A. Dyakov}
%\email[]{e-mail: sedyakov@kth.se}
\affiliation{Department of Materials and Nano Physics, School of Information and Communication Technology, KTH Royal Institute of Technology, Electrum 229, 16440 Kista, Sweden}

\author{J.~Dai}
\affiliation{Department of Materials and Nano Physics, School of Information and Communication Technology, KTH Royal Institute of Technology, Electrum 229, 16440 Kista, Sweden}

\author{M.~Yan}
\affiliation{Department of Materials and Nano Physics, School of Information and Communication Technology, KTH Royal Institute of Technology, Electrum 229, 16440 Kista, Sweden}

\author{M.~Qiu}
%\email[]{e-mail: minqiu@zju.edu.cn and min@kth.se}
\affiliation{State Key Laboratory of Modern Optical Instrumentation, Department of Optical Engineering, Zhejiang University, 310027, Hangzhou, China}
\affiliation{Department of Materials and Nano Physics, School of Information and Communication Technology, KTH Royal Institute of Technology, Electrum 229, 16440 Kista, Sweden}
\date{October 27, 2014}

\begin{abstract}
We report the concept of a {near-field} memory device based on the {radiative} bistability effect in the system of two closely separated parallel plates of \si and \va which exchange heat by thermal radiation {in vacuum}. We demonstrate that the VO$_2$ plate, having metal-insulator transition at 340 K, has two thermodynamical steady-states. One can switch between the states using an external laser impulse. We show that due to near-field photon tunneling {between the plates}, the switching time is found to be only 5\,ms which is several orders lower than in case of far field.
\end{abstract}


\maketitle %\maketitle must follow title, authors, abstract and \pacs

%%%%%%%%%%%%%%%%%%%%%%%%%%%%
{In the past decade, the proposals of thermal analogs of electronic components of integrated circuits, such as transistor \cite{li2006negative}, diode \cite{li2004thermal}, memory element \cite{wang2008thermal} and logic element \cite{wang2007thermal}, attract attention of researchers due to their potential in controlling the heat flow by phonon heat flux for information processing.} However the feasibility of this technology is limited by several fundamental constraints \cite{ben2014near}. The two main constraints are the speed of heat carriers and the Kapitza resistance also known as interfacial thermal resistance due to the small overlapping of phonon states at the interface of different elements. Another problem in taming the phonons for information processing is that a strong phonon-phonon interaction could result in non-linearity of thermal behavior of the phononic devices. Finally, the stability of the system might be reduced because of radiative losses and thermal fluctuations \cite{PhysRevLett.113.074301}. On the other hand, the use of photons as heat exchange carriers for spatially isolated objects is free from the aforementioned difficulties that opens new horizons in controlling the heat flow. The optical counterparts of diode \cite{ben2013phase, otey2010thermal}, transistor \cite{ben2014near} and memory \cite{PhysRevLett.113.074301} have been proposed.

Very recently, Kubitskyi \textit{et. al} \cite{PhysRevLett.113.074301} introduced the concept of thermal memory based on the radiative thermal bistability in the system of two parallel plates of \si and VO$_2$. The physical origin of bistability is the first-order phase transition of \va at $T_{ph} = 340$\,K \cite{qazilbash2007mott}. When the temperature of \va is smaller than $T_{ph}$ then it behaves as a uniaxial crystal with the optical axis orthogonal to its interfaces. On the other hand, when the temperature of \va is higher than $T_{ph}$, a \va plate is in its metallic phase and remains in this state for higher temperatures. In \cite{PhysRevLett.113.074301} the feasibility for thermal memory devices was demonstrated for the far-field regime. The possibility for thermal bistability in the near field regime was demonstrated by Zhu \textit{et. al} in \cite{zhu2012negative} in terms of negative differential thermal conductance for closely separated SiC plates. {We note in passing that in this work} we focus on the near field regime for \va and \si plates when separation distance between the plates is small enough for photon tunneling. In such case the heat exchange power between the plates exceeds that between two blackbodies \cite{PhysRevB.90.045414} that thereby can lead to drastic shortening of the switching time. 
\begin{figure}[b!]
\centering
\includegraphics[width=4.7cm]{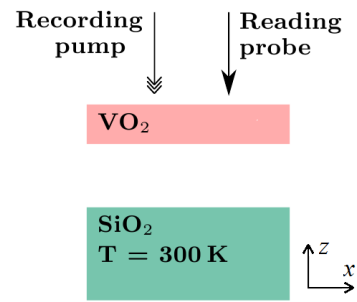}
\caption{The concept of near field thermal memory device consisting of a \si plate and a \va plate separated by thin vacuum gap. The system has two stationary states which differ in temperature and phase state of \va plate. The external impulse switches between states of \va plate.}
\label{fig0}
\end{figure}

The concept of the near field thermal memory is shown in Fig.\,\ref{fig0}. The structure consists of {a semi-infinite \si plate and a thin \va plate} separated by a vacuum gap. {The choice of \si as a material of the first plate is attributed to a strong phonon-phonon coupling between the \si and \va insulator inside the vacuum gap \cite{yang2013radiation}. The system is immersed in a thermal bath at temperature of $T_{bath}=300$\,K.} The temperature of \si plate is fixed {also} at 300\,K, while the temperature of \va is varied. We will show that although the temperature of \si plate is fixed, the system has two thermal stationary states which differ by the phase state of \va plate. In the proposed memory device, one-bit data is encoded by one of two stationary states. We will also show that due to the radiative heat exchange between plates, one can use the external laser pulse to switch between these states (see Fig.\,\ref{fig0}). %In order to read the information, stored in the \va plate, one can measure the optical transmission through the structure that will be shown to be sufficiently different for two stationary states. %{Finally, we will demonstrate the effect of near field photon tunneling on the switching time.}

%%%%%%%%%%%%%%%%%%%%%%%%%%%%

%\begin{figure}[t!]
%	\centering
%	\def\svgwidth{1\columnwidth}
%	\input{Figs/2.pdf_tex}
%	\caption{The concept of near field thermal memory device consisting of \si thick emitter plate and 50\,nm receiver plate of VO$_2$ separated by 50\,nm vacuum gap. Due do the near field interaction between emitter and receiver, the system has two stationary states which differ in temperature and phase state of \va plate. The external impulse switches between states of \va plate.}
%	\label{fig2}
%\end{figure}

\begin{figure}[b!]
\centering
\includegraphics[width=8cm]{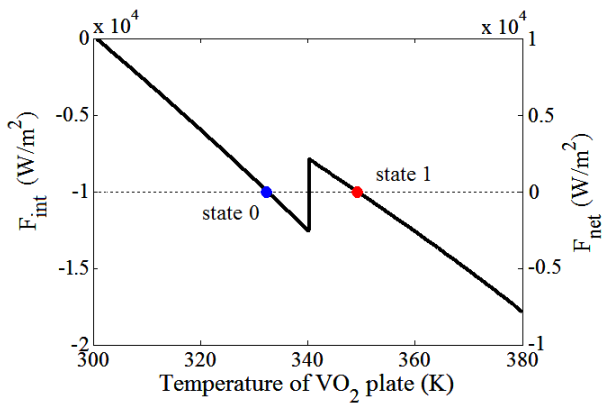}
\caption{The power of the heat transfer from \si plate to \va plate due to the difference in their temperatures, $F_{int}(T_2)$, and the net power flux $F_{net}(T_2)$ of \va plate as a function of the temperature of \va plate. The graph illustrates the negative differential thermal conductance. $h=50$\,nm, $d=50$\,nm, $F_{0}=1.9\times10^4$\,W/m$^2$, $T_{\mathrm{SiO_2}}=300$\,K.}
\label{fig1}
\end{figure}

The radiative heat exchange between plates is treated with a formalism of fluctuational electrodynamics \cite{rytov1959theory}. The energy flux of thermal radiation of $i$-th plate {at temperature $T_i$} in $z_j$ coordinate is expressed as \cite{PhysRevA.84.042102}:
\begin{equation}
\label{eq1}
F_{ij}(T_i) = \sum_{{s,p}}\int_0^{\infty}\frac{d\omega}{2\pi}\Theta(\omega, T_i)\int_0^\infty\frac{k_xdk_x}{2\pi}f_{ij}(\omega,k_x).
\end{equation}
In the above equation, index $i=1$ or 2 denotes \si and \va plates correspondingly; $z_1$ denotes any coordinate in the separation gap and $z_2$ denotes any coordinate in the upper semi-infinite vacuum; $\omega$ is the angular frequency and $k_x$ is the $x$-component of wavevector; $f_{ij}(\omega, k_x)$ is the monochromatic {transmittance} of thermal radiation at certain $k_x$; $\Theta(\omega, T_i) = \hbar\omega/\left[exp(\hbar\omega/k_BT_i)\right]$ is the mean energy of Planck oscillator and $k_B$ is Boltzmann's constant. The summation symbol in expression (\ref{eq1}) denotes the accounting for $s$- and $p$-polarizations. The monochromatic {transmittance} of thermal radiation $f_{ij}(\omega, k_x)$ is calculated in terms of the complex amplitude reflectance and transmittance of the plates. For $k_x<\omega/c$, the coefficients $f_{ij}(\omega, k_x)$ are given by the following expressions:
{
\begin{align}
\label{eq11}f_{11}&= \left(1-|r_1|^2\right) \left(1-|r_2|^2\right)|D|^{-2},\\
\label{eq21}f_{21}&= \left(1-|r_1|^2\right) \left(1-|r_2|^2-|t_2|^2\right)|D|^{-2},\\
\label{eq12}f_{12}&= \left(1-|r_1|^2\right) |t_2|^2|D|^{-2},\\
\label{eq22}f_{22}&= 1-|r_{02}|^2-\left(1-|r_{1}|^2\right)|t_{2}|^2|D|^{-2},
\end{align}
}
where $D=1-r_1r_2e^{2ik_{z0}h}$ is the Fabry-Perot like denominator and $h$ is the separation distance between plates. For $k_x>\omega/c$
\begin{align}
\label{eq1nf}f_{11}&= f_{21} = 4\mathrm{Im}(r_1)\mathrm{Im}(r_2)e^{-2|k_{z0}|h}|D|^{-2}\\
\label{eq2nf}f_{12}&= f_{22} = 0.
\end{align}
In expressions (\ref{eq11})--(\ref{eq2nf}), $r_i$ and $t_i$ are the complex amplitude reflectance and transmittance of the $i$-th plate, $r_{02}$ is the complex amplitude reflectance of the whole structure from the side of semi-infinite vacuum and $k_{z0}$ is the $z$-component of the wavevector in vacuum. Parameters $r_i$ and $t_i$ can be calculated by means of the scattering matrix method \cite{ko88, PhysRevB.90.045414, francoeur2009solution}. The Fresnel coefficients that are used in construction of the scattering matrices, accounting for anisotropy of \va plate, are described in \cite{ben2013phase, guo2014fluctuational}. {Expressions (\ref{eq11}), (\ref{eq21})  and (\ref{eq1nf}) coincide with those for the transmission coefficients of radiative heat transfer between two semi-infinite plates (see for example Ref.\,[\citenum{francoeur2011electric}]) with exception of the part $1-|r_2|^2-|t_2|^2$ in (\ref{eq21}) which accounts for the emissivity of \va plate \cite{guo2014fluctuational}. Expressions  (\ref{eq12}) and (\ref{eq22}) can be obtained from the Kirchoff's law by calculating the absorptivities of \va and \si plates when the electromagnetic wave irradiates the structure from the semi-infinite vacuum. The corresponding scattering matrix manipulations are quite straightforward and not presented here.}

At stationary state, the net energy flux emitted or received by \va plate vanishes. This is described by the following equation:
\begin{equation}
\label{eq3}
F_{net}(T_2) \equiv F_{int}(T_2)+F_{bath}+F_{ext} = 0,
\end{equation}
where $F_{int}(T_2)= F_{11}-F_{12}-F_{21}-F_{22}$ is the power of the heat transfer from \si plate to \va plate due to the difference in their temperatures. {The term $F_{bath}$ denotes the power which is absorbed in the \va plate due to thermal bath and can be calculated as $-F_{int}(T_{bath})$}. {The term $F_{ext}$ stands for the power which is absorbed in the \va plate due to some external energy source. In practice, the external energy source can be either a thermal object, a laser, or an electric heater.} In this paper, in order to control the value of $F_{net}$ we take the external energy source as a laser beam of 442\,nm wavelength, which hits the \va plate from the upper semi-infinite vacuum at normal angle of incidence. The wavelength of 442\,nm corresponds to equal extinction coefficients of \va in amorphous and metallic phases. {The power which is absorbed in \va plate is $F_{ext}=aF_0$, where $F_0$ is the laser power and $a$ is the absorption coefficient which is calculated by a standard scattering matrix formalism. For $d=50$\,nm, $h=50$\,nm $a\approx0.53$ for both states of \va plate.}

At the fixed temperature of \si plate, the zero of the function $F_{net}(T_2)$ defines the steady-state temperature of \va plate. In case of absence of the phase transitions, and when the temperature dependence of dielectric constants $\tilde{\varepsilon}_i$ is weak,  the function $F_{net}(T_2)$ has a single zero. %It means that at a given geometry of the structure and fixed emitter temperature there is only one temperature of receiver that satisfies the energy balance equation (\ref{eq3}). 
This situation is described for two parallel SiC plates in \cite{PhysRevB.90.045414}. The multiple zeros of the function $F_{net}(T_2)$ may arise due to the essential temperature dependence of dielectric constants of the material that was studied in \cite{zhu2012negative}.

Due to the phase transition, the dielectric function $\tilde{\varepsilon}_2$ of \va plate has a jump at the phase transition temperature, $T_{ph}$ \cite{PhysRevLett.17.1286, yang2013radiation}. This causes the discontinuity of the first kind of the function $F_{net}(T_2)$ at $T_2=T_{ph}$ that results in the negative differential thermal conductance between two plates (see Fig.\,\ref{fig1}). {The value of the laser power $F_{0}=1.9\times10^4$\,W/m$^2$} is chosen in such a way that the discontinuity of the function $F_{net}(T_2)$ is located near zero (see right axis scale in Fig.\,\ref{fig1}). The function $F_{net}(T_2)$ was calculated for separation distance between plates $h=50$\,nm, thickness of \va plate $d=50$\,nm and temperature of \si plate $T_{\mathrm{SiO_2}}=300$\,K. The choice of the parameter $d$ is attributed to the wishes to minimize the heat capacitance of \va plate in order to accelerate the thermal relaxation. For precisely that reason, the separation distance in turn has to be small enough to ensure efficient near-field photon tunneling from one plate to the other. In this case, the characteristic heat exchange power is of several orders of magnitude higher than in case of the far field, that at a given heat capacitance of \va plate yields in sufficient shortening of the thermal relaxation time \cite{PhysRevB.90.045414}. On the other hand, the parameters $d$ and $h$ must not be too small from the practical point of view. Thus, the \va plate has two stationary states which differ by the temperature and the phase state. {We call the situation when the $T_2 = 332.4$\,K as "state 0" and $T_2 = 349.1$\,K as "state 1". }

\begin{figure}[b!]
\centering
\includegraphics[width=7.5cm]{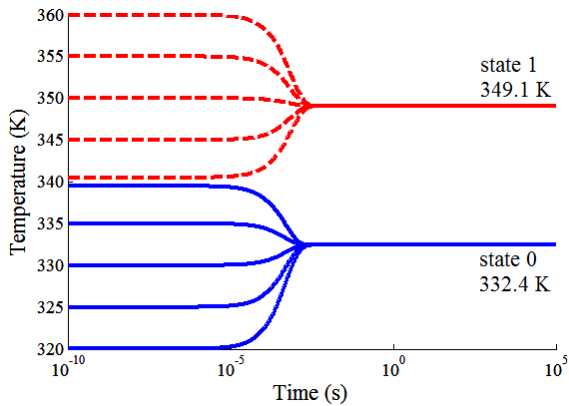}
\caption{The time evolutions of temperature of \va plate at various initial values. Solid and dashed lines denote crystalline and metallic phases of \va plate correspondingly. }
\label{fig2}
\end{figure}
In order to check whether the states 0 and 1 are stable over time let us simulate the {temporal} dynamics of the {\va plate temperature}. The energy balance equation of the radiative heat transfer dynamics is written as \cite{PhysRevB.90.045414}
\begin{equation}
\label{intdiff}
\rho c_{v}d\frac{dT_2(t)}{dt} = F_{11} - F_{12} -F_{21}(T_2)-F_{22}(T_2)+F_{bath}+F_{ext}(t),
\end{equation}
with $F_{ext}(t) = aF_0=const$.%, where $a$ is a relative portion of the external power which is absorbed in \va plate. %For $d=50$\,nm and  aforementioned parameters of the external power, $a\approx0.53$ for both states of \va plate.

In Eq. (\ref{intdiff}), $\rho = 4.6$\,g/cm$^3$ is the mass density of VO$_2$, $c_{v}$ is the mass heat capacitance at constant volume of VO$_2$. In calculations, the temperature dependence of the mass heat capacitance of \va is accounted for by the Debye model with Debye temperature of 750\,K, the molar mass of 85.92\,g/mol and the number of atoms in \va molecule of 3. In Eq. (\ref{intdiff}) the spatial temperature distribution inside the plates is assumed to be homogeneous. The solution of integro-differential equation (\ref{intdiff}) is shown in Fig.\,\ref{fig2} for the initial temperatures of \va plate, $T_2(0)$, varied from 320\,K to 360\,K. In fact, Fig.\,\ref{fig2} demonstrates the bistability in the {studied system}. Indeed, when the initial temperature of the \va plate is less than 340\,K, the steady-state temperature is 332.4\,K. On the other hand, for $T_2(0)>340$\,K, the steady-state temperature is 349.1\,K. {These temperatures correspond to those obtained from the analysis of temperature dependence of the net power flux for \va plate (see Fig.\,\ref{fig1}).} In both cases, the phase state of the \va plate does not change during the thermalization process. The time it takes for \va plate to reach stationary state is $3$\,ms and is determined by the thickness of \va plate, separation distance and the material characteristics \cite{PhysRevB.90.045414}.

\begin{figure}[b!]
\centering
\includegraphics[width=8.6cm]{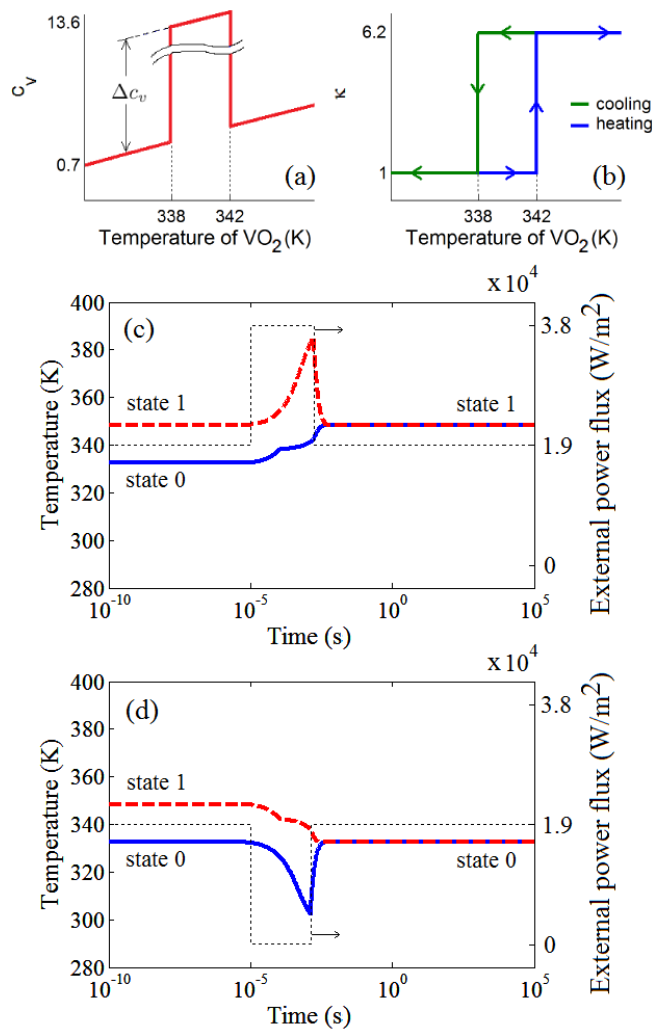}
\caption{(a) Temperature dependence of the mass heat capacitance. (b) Imaginary part of refractive index of \va at $\omega=150$\,THz near the phase transition temperature. The dynamics of switching of \va plate from (c) state 0 to state 1 and (d) from state 1 to state 0. Dotted lines show the switching impulse of external source.}
\label{fig3}
\end{figure}

So far, the time evolutions are obtained within assumption that the incident power $F_0$ is constant. When $F_0$ is no longer constant in time, the temperature of \va plate is time-dependent. Of particular interest in manipulating of \va plate temperature is the case when the function $F_{0}(t)$ is a single pulse. Let us consider the single rectangular pulse of the external power $F_{0}(t)$ with the duration of $\Delta t$ and the amplitude of $\Delta F$. Under the action of external pulse, the change of \va plate temperature is accompanied by the phase transitions. The metal-insulator transition of \va is characterized by the latent heat and hysteresis of the optical constants. In simulation of the time evolutions of \va temperature under the external pulse, the latent heat of the phase transition, $L$, is accounted for by assuming that the heat capacitance of \va in the temperature range $\{T_{ph}-\Delta T,T_{ph}+\Delta T\}$ is $c_v = c_{vo}+\Delta c_v$ (see Fig.\,\ref{fig3}a) \cite{berglund1969electronic}. Here $c_{vo}$ is the heat capacitance of \va calculated by the Debye model, and the extra heat capacitance $\Delta c_v$ is chosen in such a way that $2\Delta c_v\Delta T = L$, where $\Delta T = 2$\,K and $L = 51.49$\,J/g \cite{berglund1969electronic}. The hysteresis of optical constants is also accounted for, which is illustrated in Fig.\,\ref{fig3}b for the extinction coefficient $\kappa$ at $\omega=150$\,THz. We consider that in heating/cooling regime, the optical constants of the \va plate remain unchanged until the full amount of latent heat is absorbed.

The time evolutions of \va temperature under the positive and negative external pulses with amplitudes $\Delta F = \pm1.9\times10^4$\,W/m$^2$ and durations $\Delta t=1.7$ and 1.3\,ms are shown in Figs.\,\ref{fig3}c,d for two initial temperatures which correspond to states 0 and 1. {Note that the negative amplitude of the external pulse corresponds to decrease of the initial laser power $F_0$ to a value $F_0-|\Delta F|$.} It can be seen from Figs.\,\ref{fig3}a,b that regardless of the initial temperature of \va plate, the system {thermalizes} to state 1 after the action of a positive external pulse and to state 0 after the action of a negative external pulse. The overall time it takes to switch from state 0 to state 1 is 5\,ms and from state 1 to state 0 is 4\,ms. Thus, the time evolutions shown in Figs.\,\ref{fig3}c,d illustrate the switching procedure between states 0 and 1. Therefore, the described impulses of external power can be used for data writing to the thermal memory.
% measurements of reflection and/or transmission signal of the entire structure. The wavelength of the probe signal has to be picked from the available selection in a way to provide the maximal difference in optical constants of \va in different phases. For example, for the probe beam of 1450\,nm wavelength, the 

%by different ways including , etc. I
%\begin{figure}[b!]
%\centering
%\includegraphics[width=8.6cm]{T2D2.eps}
%\caption{The heat transmission coefficent (a) from \si plate to \va plate in its crystalline phase and (b) from \si plate to \va plate in its metallic phase calculated as $(f_{11}^{s}-f_{12}^{s}) - (f_{11}^{p}-f_{12}^{p})$, where symbols $s$ and $p$ denote different polarizations.}
%\label{fig3}
%\end{figure}

{From practical point of view, the separation distance of 50\,nm might be too small to be demonstrated. Increase of the separation distance would lead to the less efficient photon tunneling via the near field channels and as a consequence, lower transmittances of thermal radiation $f_{11}$ and $f_{21}$. This causes an increase of the switching time which, in turn, depends not only on the absolute values of fluxes $F_{ij}$, $F_{bath}$ and $F_{ext}$ but also on their relative contributions to the energy balance of \va plate. In the light of this, further investigation is needed to describe the influence of geometry, including the effect of boundaries, on the radiative bistability. In what follows we will compare the switching times in the near- and far-field regimes. In order to do that} we simulated the switching dynamics for the structure, the same as we considered so far, but with the separation distance of 5\,$\mu$m. The resulted switching time is found to be {$\sim10$\,s}. In the far field, the terms $F_{ij}$ as well as the appropriate value of external power, $F_{ext}$, in Eq.\,(\ref{intdiff}) are three orders of magnitude lower than in the near field \cite{PhysRevB.90.045414}. At a given latent heat of the phase transition and thickness of \va plate, this yields in longer switching times. {The drastic change in characteristic heat exchange rates due to the near field was also reported in \cite{tschikin2012radiative, PhysRevB.90.045414, PhysRevB.88.104307}.}

Finally, the information stored in the thermal memory can be read by measurements of temperature or electrical resistance of \va plate, or by measurements of transmission or reflection spectra. {The memory reading with resistor thermometry on the \va would probably change the size and shape of the \va layer. So that the optical transmission or reflection measurements appear to be more reasonable with such sizes.} Our scattering matrix simulations showed that the transmission coefficient of the entire structure at $\lambda=1450$\,nm for the normal incidence of the probe beam (see Fig.\ref{fig0}) equals to 0.43  in case of crystalline phase of \va plate and to 0.23 in case of metallic phase.

In conclusion, we have theoretically demonstrated the thermal radiative bistability and the memory effect in the system of two parallel plates of \si and \va separated by a thin vacuum gap. In this geometry, due to contactless near-field interaction between the plates, the driving heat exchange flux provides 5\,ms switching time which more than 3 orders of magnitude faster than in the far field. In spite of the fact that 5\,ms switching time is fairly slow, we believe that the discussed structure is a great example of radiative heat flux control {by near field} that could be useful for practical applications in information processing.

%%%%%%%%%%%%%%%%%%%%%%%%%%%%%%%%%%%%%%%%%%%%%%
%\section{Results and discussion}

%%%%%%%%%%%%%%%%%%%%%%%%%%%%%%%%%%%%%%%%%%%%%%
%\section{Conclusion}

%%%%%%%%%%%%%%%%%%%%%%%%%%%%%%%%%%%%%%%%%%%%%%

%\begin{acknowledgments}
This work is supported by the Swedish Research Council (VR) and VR's Linnaeus center in Advanced Optics and Photonics (ADOPT). M.Q. acknowledges the support by the National Natural Science Foundation of China (Grants Nos. 61275030, 61205030, and 61235007). S.D. acknowledges the Olle Eriksson Foundation for Materials Engineering for support.

%\end{acknowledgments}

%%%%%%%%%%%%%%%%%%%%%%%%%%%%%%%%%%%%%%%%%%%%%%

%%%%%%%%%%%%%%%%%%%%%%%%%%%%%%%%%%%%%%%%%%%%%%
%\bibliography{JAB_library}
%\bibliographystyle{apsrev4-1}
%

\end{document}